# Restoration of Topological Surface State by Vacuum Annealing in Magnetically Doped Topological Insulator

Jinsu Kim[1,‡], Eun-Ha Shin[2,‡], Manoj K. Sharma[3,4], Kyuwook Ihm[4], Otgonbayar Dugerjav[5], Chanyong Hwang[5], Hwangho Lee[6,7], Kyung-Tae Ko[6,7], Jae-Hoon Park[6,7,8], Miyoung Kim[2], Hanchul Kim[2*], Myung-Hwa Jung[1*]

[1]Department of Physics, Sogang University, Seoul 04107, Korea, [2]Department of Physics, Sookmyung Women's University, Seoul 04310, Korea, [3]Department of Applied Physics, Amity Institute of Applied Sciences, Amity University, Noida 201303, India, [4]Pohang Accelerator Laboratory, Pohang 37673, Korea, [5]Center for Nanometrology, Korea Research Institute of Standards and Science, Daejeon 34113, Korea, [6]Department of Physics, Pohang University of Science and Technology, Pohang 37673, Korea, [7]Max Planck POSTECH Center for Complex Phase Materials, Pohang University of Science and Technology, Pohang 37673, Korea, [8]Division of Advanced Materials Science, Pohang University of Science and Technology, Pohang 37673, Korea, [‡]These authors contributed equally to this work.

*Corresponding authors' email; mhjung@sogang.ac.kr and hanchul@sookmyung.ac.kr

The introduction of magnetic order on the surface of topological insulators in general breaks the two-dimensional character of topological surface state (TSS). Once the TSS disappears, it is improbable to restore the topological surface properties. In this report, we demonstrate that it is possible to restore the inherent TSS by ultra-high vacuum annealing. Starting from an antiferromagnetic Gd-doped $Bi_2Te_3$, that has surface state gap without TSS properties, after annealing we observed the gap closing as well as typical TSS features in physical properties. The microscopic mechanism of atomic migration and TSS restoration by annealing process is unraveled by the combination of scanning tunneling microscopy measurements and density functional theory calculations. This approach to control the surface of topological insulators and stabilize the TSS simply by vacuum annealing provides a new platform towards the exploitation of their topological properties.

Bismuth chalcogenides have attracted great attention in condensed matter physics as well as materials science because they hold enormous potential as thermoelectric materials [1,2] and topological insulators (TIs) [3-7]. Especially, TIs offer a new platform for the study of topological properties such as quantum linear magnetoresistance [8], spin-polarized two-dimensional transport [3], and Majorana fermions [9]. The most prominent property of TIs lies in gapless surface state, which is topologically protected by time-reversal symmetry [10,11]. Various research groups have reported the effect of impurities on the gapless surface state with linear energy dispersion [12-16]. Nonmagnetic impurities such as $NO_2$, Ca, and Rb give almost no effect on the linear energy dispersion but play an important role in tuning the Fermi level [12,13], preserving the topological surface state (TSS). The electrical transport of spin-momentum-locked TSS is believed to be robust in the presence of nonmagnetic perturbations. However, TSS can be broken under magnetic perturbations within the proximity to magnetic materials. Local ferromagnetic phases or magnetic impurities such as Mn and Fe on the surface induce an energy gap at the Dirac point [4,7,11], which modifies the energy dispersion relation of TSS. Moreover, since the transition metals of Mn and Fe are mostly divalent and replace the trivalent Bi ions, such magnetic impurities of transition metals act as hole dopants in TIs and affect the magneto-transport properties.

In order to tune only the magnetic properties of TIs, trivalent rare-earth metals are suitable candidates. In our previous report on Gd-substituted $Bi_2Te_3$, $Gd_xBi_{2-x}Te_3$, non-trivial topological features were observed near the magnetic phase transition at $x_c = 0.09$ from paramagnetic (PM) to antiferromagnetic (AFM) phase [17]. The experimental observations near $x_c$ are high electrical resistivity, large linear magnetoresistance, and highly anisotropic Shubnikov-de Haas (SdH) oscillations with two-dimensional (2D) TSS origin. On the other hand, the samples of $x > x_c$ show AFM ordering and normal metallic behavior without TSS features. Such absence of TSS properties may come from magnetic order or impurity scattering, which is not clear yet. Most of previous works focus on the gap opening of surface band by doping magnetic impurities into TIs. One meaningful question is whether the TSS nature once destroyed by the magnetic impurities can be restored by controlling the lattice disorder and magnetic coupling strength.

Here we demonstrate that the restoration of TSS in $Gd_{0.15}Bi_{1.85}Te_3$ is achieved by ultrahigh vacuum annealing. We investigate the magnetization, magneto-transport, band structure, and surface morphology before and after vacuum annealing, in combination of theoretical calculation. We find that Gd atoms migrate to top surface after vacuum annealing and leave Bi atoms on the top layer. The rearrangement of magnetic Gd atoms provokes a magnetic transition from AFM to PM phase, and the topological 2D transport features such as large linear magnetoresistance, SdH oscillations with 1/2 offset, and multiband Hall behavior are observed. These results are consistent with angle-resolved photoemission spectroscopy (ARPES) measurements, which show gap closing after annealing. The combination of scanning tunneling microscopy (STM) experiments and density functional theory (DFT) calculations reveal the microscopic picture of atomic rearrangements. By engineering the defect sites via vacuum annealing, we could achieve the tuning of the Fermi level and thereby the restoration of TSS in TIs.

## Results and discussion

### Atomic migration after vacuum annealing.

We have first examined the chemical state of elements of $Gd_xBi_{1-x}Te_3$ samples by near edge x-ray absorption fine structure (NEXAFS) and x-ray photoelectron spectroscopy (XPS) measurements, which especially give information about chemical bonding state on the surface. The spectra were taken for as-grown samples (called "as-grown") and *in-situ* annealed samples in an ultra-high vacuum (called "annealed"). Figure 1a shows Gd N-edge NEXAFS spectra of $Gd_xBi_{1-x}Te_3$ ($x$ = 0.06, 0.09, 0.15, and 0.20) samples. The sharp peak at 144.8 eV is assigned to the 4*d*-4*f* transition of Gd [18]. The spectral intensity of this Gd peak increases with increasing the Gd concentration, $x$, and it is enormously enhanced for all the samples after annealing. In the inset of Fig. 1a, a typical intensity enhancement after annealing is clearly shown in a magnified scale. This observation implies that the Gd atoms are more populated at the surface after the vacuum annealing, possibly due to migration from the bulk. Similarly, the chemical state of Bi element is determined from the XPS spectra. Figures

1b and 1c display the Bi $4f$ core-level XPS spectra for $x = 0.15$ before and after annealing. The Bi $4f$ spectra show a simple spin-orbit doublet structure. The two main peaks around 157.0 and 162.3 eV correspond to the binding energy of Bi $4f_{7/2}$ and Bi $4f_{5/2}$, respectively. Each core-level spectrum was fitted with two components using a least-squares fitting and mixed Gaussian-Lorentzian functions. An additional asymmetry parameter was used for metallic components in higher binding energy side [19,20]. In Fig. 1b, the Bi $4f$ spectrum for as-grown sample is fully fitted only with bound Bi states after subtraction of a normal state background, in good agreement with the spectra obtained from pristine $Bi_2Te_3$. After annealing in Fig. 1c, the spectrum is resolved into two chemical states of bound Bi and unbound Bi. The unbound Bi peaks have a slight shift (~ 0.2 eV) towards the lower binding side from the bound Bi peaks. The area ratio of the bound Bi is relatively reduced with the increased portion of unbound Bi, indicating that the Bi ion breaks the bonds with the Te ions and resides at the surface as metallic Bi states. On the other hand, the XPS spectra of Te $3d$ core level did not show any significant difference after annealing (Supplementary Fig. 3), suggesting not only the intact chemical state of Te but also no oxidation in the vacuum chamber during the XPS measurement. From the combined NEXAFS and XPS investigations, we speculate that the Gd ions migrate from the bulk to the surface during the vacuum annealing, and that the surface Bi atoms attain increased metallic nature.

**Changes of physical properties after vacuum annealing.**

We examined the effect of annealing on the magnetic and magneto-transport properties such as the magnetic susceptibility $\chi$, the magnetization $M$, the magnetoresistance $MR$, and the Hall resistivity $\rho_{xy}$. According to our previous studies on $Gd_xBi_{1-x}Te_3$ in ref. [17], there is a magnetic transition from PM to AFM at the critical Gd composition of $x_c = 0.09$. Among them, we chose an AFM crystal of $x = 0.15$ above $x_c$ because the Gd atoms are more probable to relocate by annealing in Gd-richer samples and thereby the AFM phase can be changed. As seen in Fig. 2a, the $\chi(T)$ curve of as-grown sample shows an AFM feature with the Néel temperature, $T_N = 11.5$ K, which well agrees with the previous data [17]. Interestingly, this AFM fingerprint disappears in two annealed samples; one is the annealed sample and the other is the sample left after the removal of the surface part by cleaving the annealed sample (called "c-annealed"). There is no significant difference of $\chi(T)$ behavior between annealed and c-annealed samples except the magnitude. The magnitude for the c-annealed sample is lower than that for the annealed sample. Reminding that more Gd atoms are present on the surface after annealing, this result is due to the reduction of magnetic moments as a result of removal of the Gd-rich surface part. Since the magnetic measurement is a bulk property, we can evaluate the physical parameters with the bulk origin. From the Curie-Weiss fit in the annealed samples, we obtain the Weiss temperature of $\theta_p$ = -0.67 K, which is almost identical value ($\theta_p$ = -0.38 K) for the critical composition of $x_c$ = 0.09 in ref. [17]. Another fitted parameter of effective magnetic moment gives an important information on the Gd concentration. The c-annealed sample shows the magnetic moment similar to that of the as-grown sample with $x$ = 0.10 (~ $x_c$). In a similar way, the $M(H)$ data can be understood. The $M(H)$ curve in Fig. 2b tends to increase before annealing, while it tends to saturate after annealing. Our previous $M(H)$ data revealed the diverging signal for AFM ($x > x_c$) and the saturating behavior for PM ($x \leq x_c$). The sample with $x$ = 0.15 ($> x_c$) seems to turn into a sample with $x$ = 0.10 (~ $x_c$) after annealing. Thus, we expect the non-trivial topological features after annealing, as observed at $x_c$ in the previous work [17], so that we check this scenario by measuring the magneto-transport properties.

The magnetoresistance (MR) ratio as a function of applied magnetic field is plotted in Fig. 2c. The MR value at 7 T is 360% for as-grown sample, and after annealing it enormously increases up to around 1400%, close to the MR value for the critical composition $x_c$. Such large and linear MR behavior is considered as a signature of the 2D properties of TSS electrons. This tendency well agrees with the magnetic data shown in Figs. 2a and 2b. The Gd concentration is reduced from $x$ = 0.15 (that is AFM mainly having bulk properties) to $x$ = 0.10 (that is PM exhibiting 2D properties of TSS) by annealing. Another evidence on the existence of TSS after annealing can be found in the Hall measurements. In Fig. 2d, the Hall curve measured with the c-annealed sample shows a weak modulated signal at low fields, so-called bent Hall effect, which occurs via two parallel conducting channels of bulk and surface carriers [21-24]. This result is in contrast with the linear behavior of the as-grown sample. Further evidence of the TSS electrons can be found in quantum oscillations. This measurement gives critical insight in the information of Fermi surface. For the c-annealed sample, the MR curve in Fig. 2c shows SdH oscillations at high magnetic fields. Since the resistivity tensor is an inverse of the conductivity tensor, we took the conductivity, $\sigma_{xx} = \rho_{xx} / (\rho_{xx}^2 + \rho_{xy}^2)$ with respect to magnetic field and plotted it against the inverse of magnetic field. After subtracting background signals with polynomial form, the lower inset of Fig. 2d clearly shows SdH oscillation above 0.15 T$^{-1}$. By identifying the minima to integer $n$ and the maxima to $n + 1/2$, we constructed the Landau level (LL) fan diagram in the upper inset of Fig. 2d. The linear fit of the LL fan diagram extrapolates to 0.62 ± 0.04 on the $n$-index axis, which is close to the value 1/2 *expected for the Dirac fermions*. The small deviation from 1/2 is reasonable because the bulk electrons also contribute to the conduction, as discussed above in the bent Hall curve.

**Restoration of TSS band structure in annealed sample.**

The main conclusion in this study is that the TSS can be restored simply by vacuum annealing. In order to directly examine how the TSS is evolved after annealing, we performed the ARPES experiments. In the as-grown sample, we observe a gap opening at the Dirac point with the gap size of 60 meV, as shown in Fig. 2e. Such a gap opening and the AFM nature upon Gd substitution suggest that the band topology and the magnetic phase is possibly interconnected, thereby losing the 2D properties of TSS. After vacuum annealing, intriguingly, the surface energy gap disappears and the linear energy dispersion of TSS is restored as shown in Fig. 2f. The electronic band structure after annealing becomes the same as the pristine Bi$_2$Te$_3$ sample, except a slight downward shift of the Fermi level in energy. This observation is consistent with the results discussed above. The restoration of TSS after annealing closes the energy gap, revealing the non-trivial topological properties such as large linear magnetoresistance, SdH oscillations with 1/2 offset, and multiband Hall behavior.

**Microscopic understanding of annealing effect by STM measurements and DFT calculations.**

To understand the Gd out-diffusion and the restoration of TSS at the atomic scale, we performed

STM measurements and DFT calculations. Figure 3a is a typical empty-state STM images of the (111) cleavage surface of as-grown samples, where Te atoms on the top are imaged as bright protrusions. Native defects, observed in pristine $Bi_2Te_3$ sample, like $Te_{Bi}$-type antisites and Bi vacancies are commonly observed in all Gd-containing samples (A, B, and C defects in Supplementary Fig. 4). Upon Gd substitution, the density of the native defects is significantly reduced and two new Gd-induced defects, α and β, are observed. The defect α appears as a triangular dark depression occupying the top surface Te site, and the defect β appears as a three-leaf-clover-shaped dark depression centered at the hcp site which is on top of the second atomic layer (Bi1). The annealed sample reveals quite different features from the as-grown sample, as shown in Figs. 3b and 3c. Compared to Fig. 3a, the α defects almost disappear, the β defects increase, and the γ defects are newly formed. The γ defect appears as a bright protrusion at the surface Te site in both filled- and empty-state images.

Possible defects induced by Gd substitution are adsorption on the surface ($Gd_{ad}$), intercalation in the van der Waals gap ($Gd_I$), and substitution for Bi ($Gd_{Bi}$) or Te ($Gd_{Te}$). From the top view of the crystal, there are three different sites; top surface Te site (top), on-top of Bi1 site (hcp), and the hollow site (fcc), as indicated in Fig. 3d. The atomic layers in a quintuple layer (QL) of $Bi_2Te_3$ are denoted by Te1-Bi1-Te2-Bi2-Te3 from the top surface. Our DFT total energy calculations (see Supplementary Table 1) show that the substitutional Gd at Bi sites ($Gd_{Bi1}$ and $Gd_{Bi2}$) are the most stable configurations with the energy difference as small as 5 meV. The next stable structure is a pair of $Gd_{Bi2}$ and $Bi_I$ ($Gd_{Bi2}$-$Bi_I$), where $Bi_I$ is the interstitial Bi in the van der Waals gap, formed via the kick-out of Bi by $Gd_{Bi2}$. The $Gd_{Bi2}$-$Bi_I$ pair is followed by the interstitial Gd ($Gd_I$) in the van der Waals gap, but its formation energy is larger than that of $Gd_{Bi2}$-$Bi_I$ by 0.18 eV. These results are in accordance with the previous reports of preferential substitution of Bi by transition-metal elements such as Cr, Mn, and Fe [25]. Gd adatoms on the surface is unlikely to exist because the formation energy ranging from 4.84-7.30 eV is much larger than those of substitutional Gd defects.

In order to identify the Gd-related defects (α, β, and γ), we simulated STM images for all possible defect types. In the enlarged STM images of Fig. 3e, the α defect appear as a dimmer protrusion taking a single Te site in the filled state and a single Te depression in the empty state. The simulated STM images of the $Gd_{Bi2}$-$Bi_I$ pair are in good agreement with experiments. As for the β defect, the STM images in Fig. 3f show a triangular bright protrusion in the filled state and a three-leaf-clover-shaped dark depression in the empty state. Such experimental features are well reproduced by the simulation of $Gd_{Bi1}$. Similar STM features have been reported for Fe-doped $Bi_2Te_3$ [26], Mn-doped $Bi_2Te_3$ [14], Cr-doped $Sb_2Te_3$ [27], and Ca-doped $Bi_2Se_3$ [28], and were attributed to the impurity elements substituted for Bi1 sites. In Fig. 3g, the γ defects appear as a bright protrusion on the top Te sites in both filled- and empty-state images, and are identified to be $Bi_{Te1}$ antisite defects based on the good agreement between experimental and simulated images. It is noticeable that the $Bi_{Te}$-type antisite defect is reported to have low formation energy under Bi-rich condition [29-31], where a clover-shaped protrusion is assigned to the $Bi_{Te3}$ antisite. The relative energetic stability of $Bi_{Te1}$ and $Bi_{Te3}$ in a slab geometry has not been studied yet, and our calculations show that $Bi_{Te3}$ is energetically more favorable than $Bi_{Te1}$ by 0.2 eV. This energetic favor of $Bi_{Te3}$ is in apparent contradiction to the sole observation of $Bi_{Te1}$ in the annealed sample, which suggests that the formation of the γ defect is driven by kinetics rather than by thermodynamics.

In short, the Gd-related defects, α, β, and γ are identified to be the $Gd_{Bi2}$-$Bi_I$ pair, the substitutional $Gd_{Bi1}$, and the $Bi_{Te1}$ antisite, respectively. The appearance of α and β defects in the as-grown sample is due to the relatively low formation energy of substitutional $Gd_{Bi}$. After annealing, the disappearance of $Gd_{Bi2}$-$Bi_I$, the increase of $Gd_{Bi1}$, and the appearance of $Bi_{Te1}$ can be schematically speculated, as depicted in Fig. 4. The volatile Te ions are evaporated upon heating, mostly from the top surface layer, generating $V_{Te1}$ [step 1]. Then, the region near the surface becomes effectively Bi-rich (or Te-poor) phase, compared to the Te-rich condition of the as-grown sample. This change is in commensurate with the disappearance of the native defects (A, B, and C) which are stable in Te-rich condition. The Te vacancies can be occupied by the adjacent Bi atoms to form the $Bi_{Te1}$-$V_{Bi}$ pair [step 2]. The subsequent liberation of $V_{Bi}$ can result in the creation of $Bi_{Te1}$, the γ defect [step 3]. Then, the liberated

$V_{Bi}$ can be occupied by the out-diffused Gd to form a $Gd_{Bi1}$, resulting in the increase of β defect. Alternatively, the liberated $V_{Bi}$ can be occupied by the interstitial Bi in the van der Waals gap, $Bi_I$ of the $Gd_{Bi2}$-$Bi_I$ pair. This causes the consumption of the $Gd_{Bi2}$-$Bi_I$ pairs, removing the α defects. The isolated $Gd_{Bi2}$ defects after consuming the $Bi_I$ atoms are invisible in the STM images (see Supplementary Fig. 5). Such atomic rearrangements result in a Gd-rich regime near the surface and the relatively dilute Gd bulk phase, which leads to an effective reduction of magnetic moment and the restoration of TSS.

In addition, the core-level shift calculations within the initial state theory show that the binding energy (BE) of Bi is shifted by -1.1 and -0.3 eV for $Bi_{Te1}$ antisite (γ defect) and its first nearest neighbor Bi atom, respectively. This is in good agreement with the observed BE shift towards lower binding energy side, as shown in Fig. 1c. For $Bi_I$ in the $Gd_{Bi2}$-$Bi_I$ pair (α defect), the calculated core-level shift is as small as 0.2 eV towards higher energy side. As for $Gd_{Bi1}$ antisite (β defect), there is no appreciable core-level shift of neighboring Bi atom (0.01 eV towards higher energy side). Our DFT calculations enables the identification of defects formed upon annealing and enlightens the possible atomistic processes. There are two main atomic rearrangements; one is the out-diffusion of Gd to the surface and the other is the formation of unbound Bi on the surface in the form of $Bi_{Te1}$. These results not only agree with our NEXAFS, XPS, and STM measurements, but also explain the magnetic and magneto-transport data.

# Methods

**Single crystal growth and structural characterization.** The single crystals of $Gd_xBi_{2-x}Te_3$ were synthesized by melting method in a vertical tube furnace with local temperature gradient. A mixture of Bi (99.999%), Te (99.9999%), and Gd (99.99%) with the stoichiometric ratio was put into a cleaned quartz tube with 5% excessive Te and sealed in vacuum. The Te excess acts to compensate the Te loss caused by its high vapor pressure, and the Te-rich condition tends to generate native $Te_{Bi}$-type antisite defects, leading to *n*-type charge carriers. The mixture in the evacuated quartz ampoule was melted at 800°C and annealed at 550°C for 3 days. The obtained ingots were well cleaved with a mirror-like surface perpendicular to the *c* axis. Single-crystal XRD data show (003) family reflections for all samples without any impurity and secondary phases (see Supplementary Fig. 1).

**XPS and ARPES measurements.** The samples ($Gd_xBi_{2-x}Te_3$) were loaded and *in-situ* cleaved in an ultra-high vacuum (UHV, base pressure: $1 \times 10^{-10}$ Torr) for XPS, NEXAFS and ARPES measurements. The *in-situ* annealing (250 °C for 2 h), and characterization of the samples by XPS and NEXAFS spectra (4D beamline) and ARPES (4A1 beamline) were performed at Pohang Accelerator Laboratory (PAL). The in-situ ARPES measurements were carried out at 20 K by using a Scienta SES2002 electron energy analyzer. XPS and Total Electron Yield (TEY) NEXAFS spectra were recorded at 45° angle of x-ray incidence at sample surface. The acquired spectra were normalized to the incident photon flux and calibrated using Au 4*f* and C 1*s* core level peaks. The Shirley-Sherwood method was used for subtraction of the secondary electron background.

**STM measurements.** The scanning tunneling microscopy (STM) measurements were carried out at 300 K in an ultrahigh vacuum with a base pressure of $2 \times 10^{-10}$ Torr, using a home-built STM with commercial RHK R9 controller. The STM images were acquired in the constant-current mode using electrochemically etched polycrystalline W tips. All samples were cleaved in an ultrahigh vacuum by using scotch tape.

**Calculations.** The first-principles calculations were performed by using a plane-wave basis, projector augmented wave potentials [32] as implemented in Vienna *ab initio* simulation package (VASP) [33]. The kinetic energy cutoff for the plane waves was 300 eV. The exchange-correlation interaction of electron was treated by the Perdew-Burke-Ernzerhof implementation of the generalized gradient approximation [34] in conjunction with the on-site Coulomb repulsion ($U = 7$ eV) for Gd $f$-orbital [35]. To describe long-ranged vdW interaction, DFT-D3 was applied [36]. We employed repeated slab geometry including 20Å-thick vacuum. Monkhorst-pack mesh of k-points (2×2×1) was used for sampling the two-dimensional Brillouin Zone [37]. For the energetics, spin-polarized calculation was carried out using a 4×4 surface supercell containing 3QL slab. The atomic positions were relaxed until Hellmann-Feynman forces are smaller than 0.01 eV/Å$^{-1}$. For the STM image simulations and core-level shift calculations [38], a 6×6 surface supercell with 1QL is used and the spin-orbit coupling is explicitly treated [39]. (For the configurations with interstitial defects, a 2QL slab was employed.) The constant-current STM images were simulated within Tersoff-Hamann approximation [40, 41].

**Magnetic and electrical properties measurements.** The magnetic and transport properties were measured in a temperature range of 2 K to 300 K with the superconducting quantum interference device-vibrating sample magnetometer (SQUID-VSM) up to 7 T. The magnetization was measured at 2 K and as sweeping magnetic field parallel to the cleaved surface of single crystals. In the same configuration, the magnetic susceptibility was measured in a magnetic field of 1 T. All transport data were taken using a dc four-probe method with conducting Pt wires. Magnetoresistance and Hall curves were obtained from the perpendicular configuration of the magnetic field perpendicular to the cleaved surface of single crystals.

**Acknowledgements**

This work was supported by the National Research Foundation of Korea (NRF) grant funded by the Korea government (MEST) (No.2014R1A2A1A1105401, 2017R1A2B3007918, 2012M3C1A6035684, 2015R1A2A2A01005564, 2017R1A2B4012972, 2015M2A2A6A01045343, and 2016K1A4A4A01922028). The calculations were supported by Supercomputing Center/Korea Institute of Science and Technology Information with supercomputing resources including technical support (No. KSC-2015-C2-049). Experiments at PLS-II were supported by MSIP-R. O., Korea.

**Author contributions**

J.K., M.K.S., K.I., O.D., C.H., H.L., K.-T.K., J.-H.P., and M.-H.J. contributed the experimental part, and E.-H.S., M.K., and H.K. contributed to the theoretical part. J.K., E.-H.S., H.K., and M.-H.J. wrote the main manuscript text. All authors reviewed the manuscript.

**Additional information**

Supplementary information

**Competing financial interests**

The authors declare no competing financial interests.

**Figures**

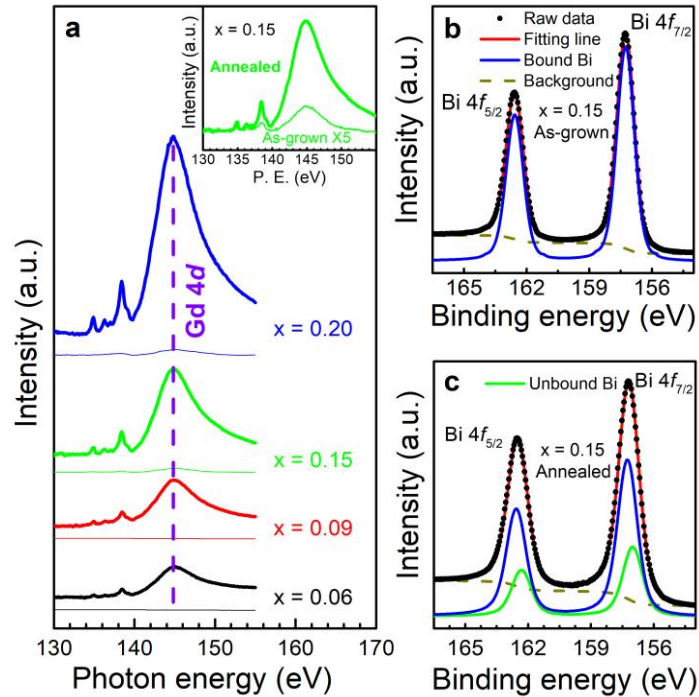

**Figure 1 Near edge x-ray absorption fine structure (NEXAFS) and x-ray photoelectron spectroscopy (XPS) of $Gd_xBi_{2-x}Te_3$.** (a) Gd N-edge NEXAFS spectra taken for as-grown and annealed samples with various $x$ (= 0.06, 0.09, 0.15, and 0.20). The thin and thick lines correspond to the spectra for as-grown and annealed samples, respectively. The inset shows a typical comparison of the NEXAFS spectrum at $x = 0.15$ before and after annealing. The data of as-grown sample have been multiplied by 5. (b) and (c) Bi 4$f$ core-level XPS spectra of $x = 0.15$ taken for as-grown and annealed samples, respectively. Before annealing, the data are fitted using two peaks for Bi $4f_{7/2}$ and Bi $4f_{5/2}$ with binding energies of 157.0 and 162.3 eV, in addition to a normal-state background signal (dashed line). After annealing, the data are resolved into two chemical states of bound Bi (blue line) and unbound Bi (green line) with a slight energy shift, about 0.2 eV.

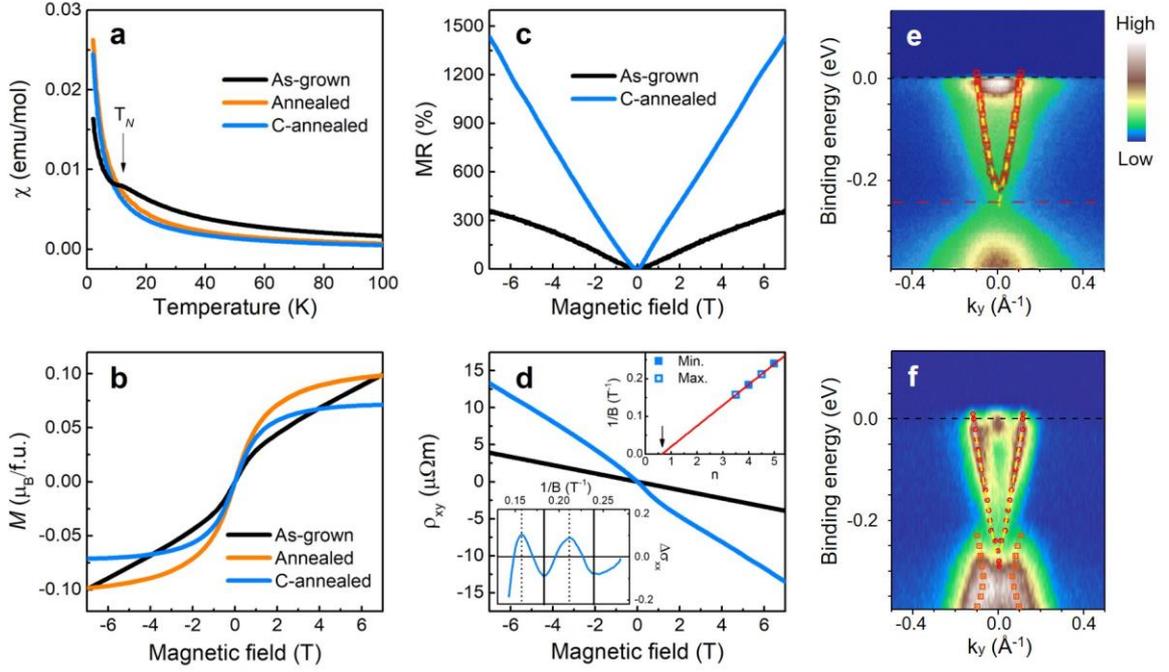

**Figure 2 Changes in physical properties and band structure upon annealing.** (a) Magnetic susceptibility as a function of temperature, $\chi(T)$ measured at 1 T. For the as-grown sample, the arrow represents the Néel temperature $T_N$ = 11.5 K. Both annealed and c-annealed samples show no such a feature of antiferromagnetic transition. (b) Magnetization as a function of the applied magnetic field, $M(H)$ measured at 2 K. A diverging signal for the as-grown sample changes into saturating behavior for both annealed and c-annealed samples. (c) Magnetoresistance, MR ratio measured at 2 K. The c-annealed sample shows large and linear MR behavior. (d) Hall resistivity, $\rho_{xy}$ measured at 2 K. The c-annealed sample shows a bent curve at low fields (blue line), distinguished from the linear behavior of the as-grown sample (black line). The lower inset represents the quantum oscillations of electrical conductivity $\Delta\sigma_{xx}$, from which the Landau level fan diagram is plotted in the upper inset. The arrow indicates the phase offset, close to 1/2 expected for the Dirac fermions. (e) and (f) Angle-resolved photoemission spectroscopy (ARPES) spectra of $x$ = 0.15 taken for as-grown and c-annealed single crystals, respectively, measured at 20 K. The as-grown sample shows an energy gap of 60 meV at the Dirac point, while after annealing the energy gap is closed and the surface states are restored.

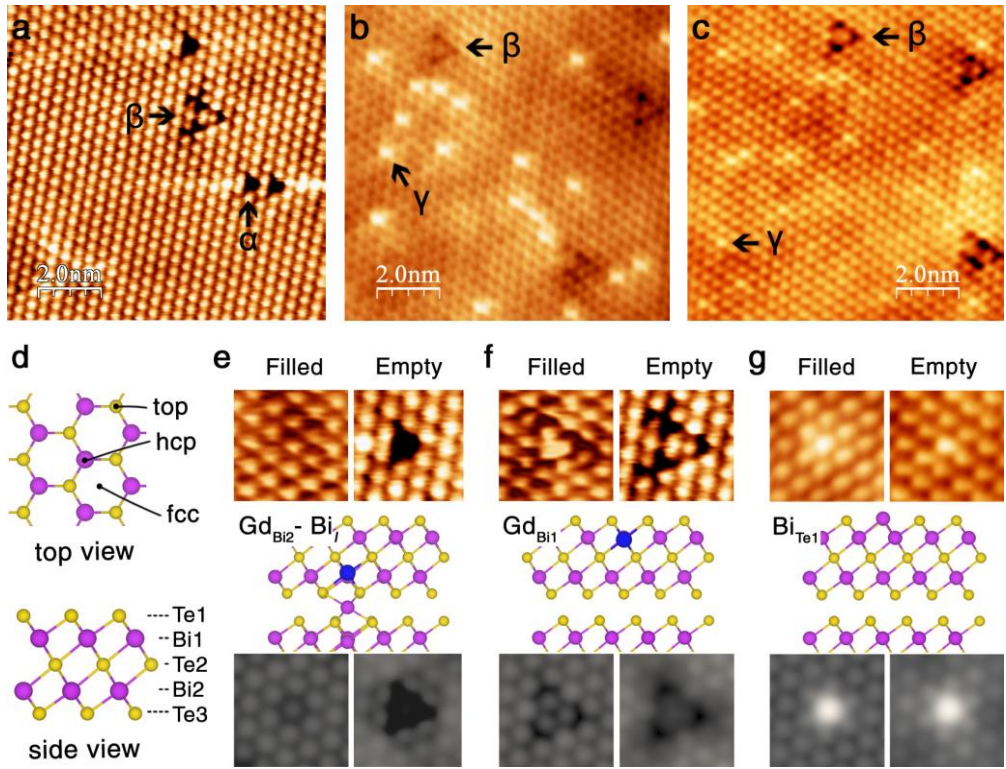

**Figure 3 Experimental and simulated STM images of x = 0.15 before and after annealing.** (a) Empty-state scanning tunneling microscopy (STM) image for as-grown single crystal. The defects α and β are related with the Gd substitution. (b) and (c) Filled- and empty-state STM images after annealing. Notice that the α defects almost disappear, the β defects increase, and the γ defects are newly formed. (d) Schematic diagram of the atomic structure of pristine $Bi_2Te_3$: top view (upper) and side view (lower). The pink and yellow circles represent Bi and Te atoms, respectively. (e)-(g) Experimental high-resolution STM images (upper), atomic structures (middle), and simulated constant-current STM images (lower). (e) α defect: $Gd_{Bi2}$-$I_{Bi}$, (f) β defect: $Gd_{Bi1}$, and (g) γ defect: $Bi_{Te1}$. The simulated STM images are in good agreement with experiments. The blue circle represents Gd atom. The filled- and empty-state images are measured at $V_S$ = -0.35 and +0.30 V in as-grown sample and $V_S$ = -0.65 and +0.35 V in annealed sample, respectively, and simulated at $V_S$ = -0.6 and +0.6 V, respectively.

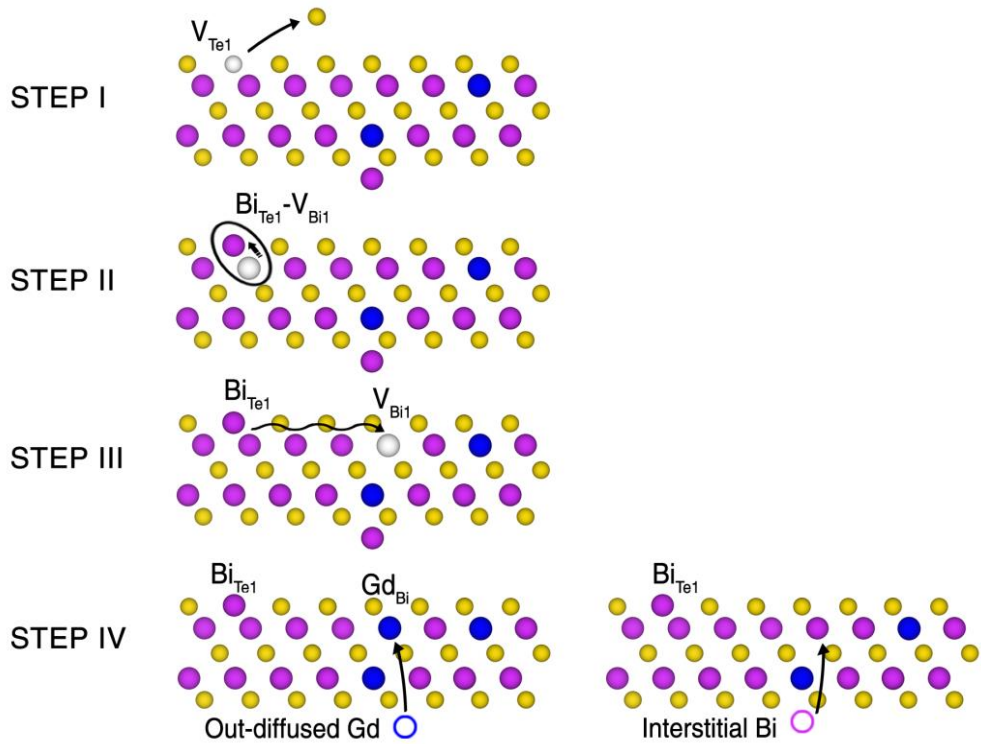

**Figure 4 Schematics of a possible atomic rearrangement process during vacuum annealing.** In step I, $V_{Te1}$ is formed by the evaporation of surface Te atoms during heating. In step II, a $Bi_{Te1}$-$V_{Bi1}$ pair is formed by the migration of Bi atom from an adjacent Bi site. In step III, $V_{Bi1}$ is liberated to form isolated $Bi_{Te1}$. In step IV, $V_{Bi1}$ is filled either by out-diffused Gd to generate $Gd_{Bi1}$ (β defect; left) or by the interstitial Bi of a $Gd_{Bi2}$-$Bi_I$ pair to generate $Bi_{Te1}$ (γ defect; right). Notice that the final structure of the right process of step IV, $Gd_{Bi2}$, is indistinguishable from the pristine surface (see Supplementary Fig. 5)

# Supplementary Information

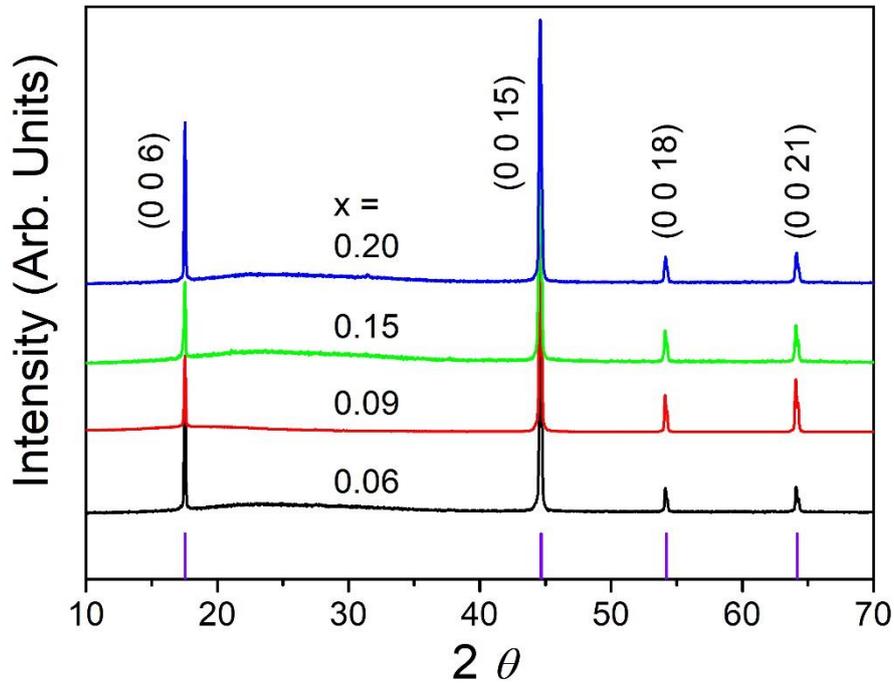

**Supplementary Figure 1. Single-crystal x-ray diffraction (XRD) patterns of $Gd_xBi_{2-x}Te_3$.** The diffraction patterns possess the (003) family peaks without any impurity peaks, and the peak positions are in good agreement with the reference of $Bi_2Te_3$ (violet column, ICSD code. 15753). This result indicates that the $Gd_xBi_{2-x}Te_3$ crystallizes in a single crystal having the same structure as $Bi_2Te_3$. The lattice parameters are $a$ = 4.39 ± 0.01 Å and $c$ = 30.48 ± 0.04 Å, which are not affected by Gd substitution because of the small difference in ionic sizes between Bi and Gd.

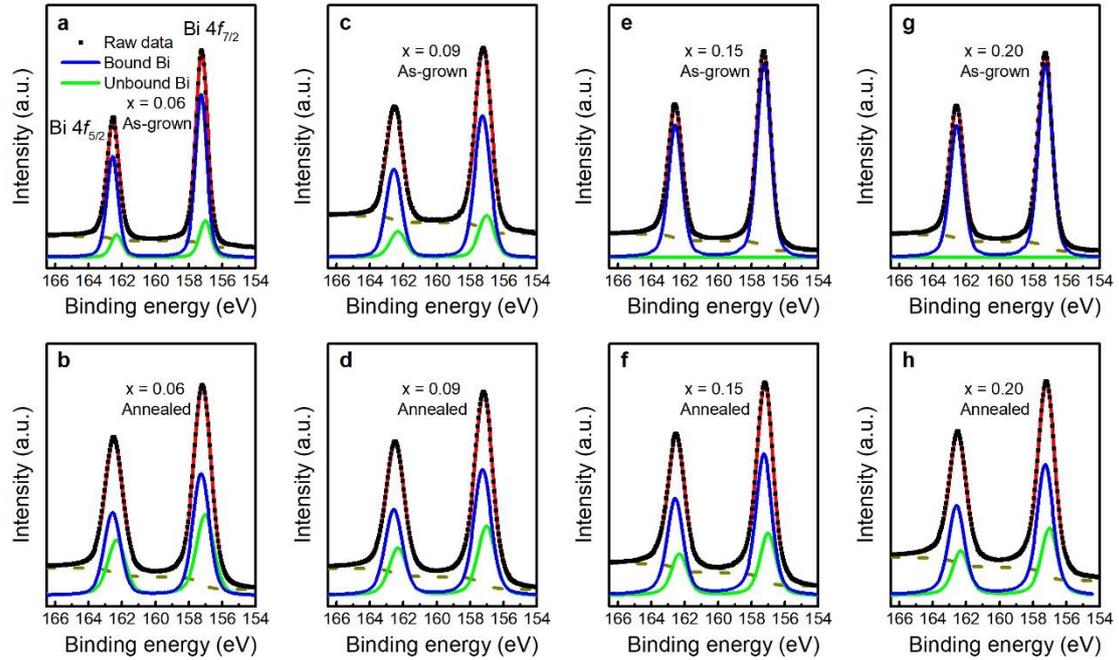

**Supplementary Figure 2. Bi 4$f$ core-level x-ray photoelectron spectroscopy (XPS) spectra taken for as-grown and annealed Gd$_x$Bi$_{2-x}$Te$_3$ single crystals.** (a), (c), (e), and (g) ((b), (d), (f), and (h)) Bi 4$f$ core-level XPS spectra of as-grown (annealed) samples for $x$ = 0.06, 0.09, 0.15, and 0.20, respectively. Two main peaks around 157.0 and 162.3 eV correspond to the binding energy of Bi 4$f_{7/2}$ and Bi 4$f_{5/2}$, respectively. Each spectrum was fitted in the same way as described in the main text. The background signal is represented by a dashed line. The enhancement of unbound Bi components after annealing implies that the Bi ion breaks the bonds with the Te ions and is present on the surface as a metallic Bi state.

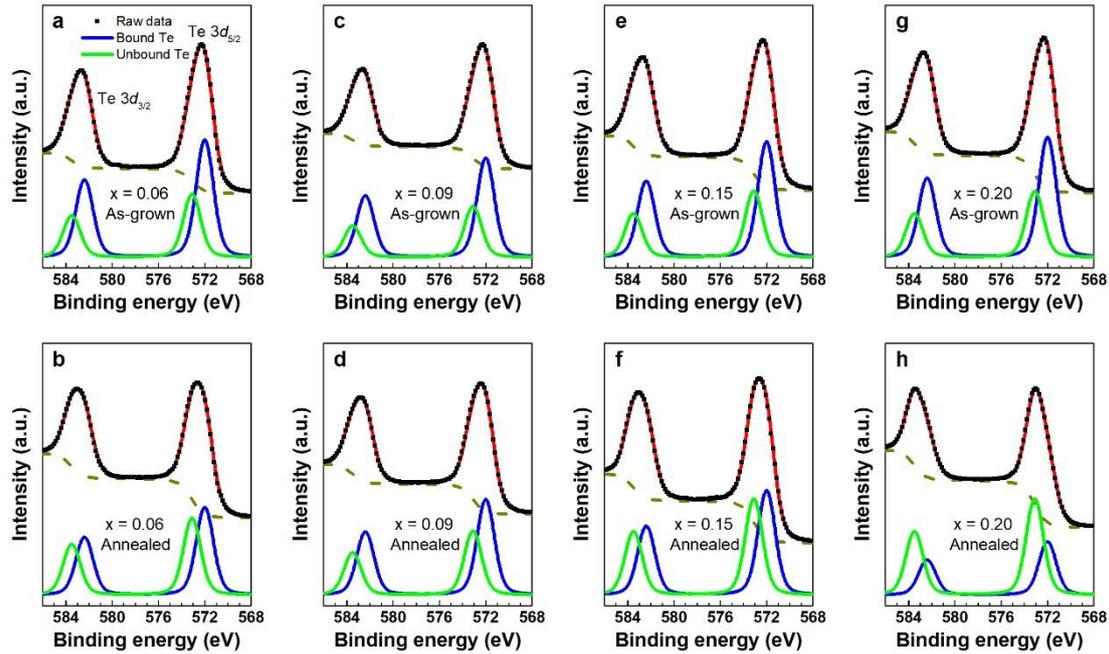

**Supplementary Figure 3. Te 3$d$ core-level x-ray photoelectron spectroscopy (XPS) spectra taken for as-grown and annealed Gd$_x$Bi$_{2-x}$Te$_3$ single crystals.** (a), (c), (e), and (g) ((b), (d), (f), and (h)) Te 3$d$ core-level XPS spectra of as-grown (annealed) samples for $x$ = 0.06, 0.09, 0.15, and 0.20, respectively. Two main peaks around 572.4 and 582.8 eV correspond to the binding energy of Te 3$d_{5/2}$ and Te 3$d_{3/2}$, respectively. The spectra were fitted in the same way as the Bi 4$f$ core-level analysis. The presence of unbound Te component in all the as-grown samples is due to the Te-rich crystal growth condition, which tends to generate the Te$_{Bi}$-type antisite defects. After annealing, deconvoluted results show that the unbound (bound) Te component is slightly increasing (decreasing), although the difference of area ratio for deconvoluted peaks is negligible between as-grown and annealed samples. Such small difference results from the volatile nature of Te atom during the annealing process. Note that oxygen-related peaks (i.e., Te-O bonds around 576.1 or 577.3 eV) are not detected [1].

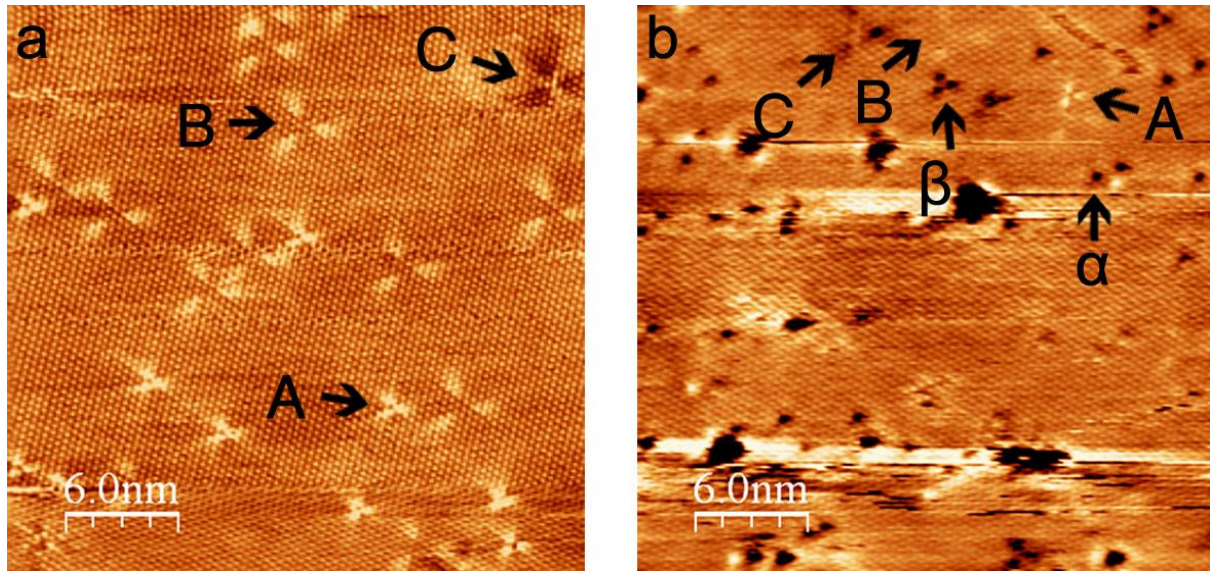

**Supplementary Figure 4. Empty-state STM image of pristine and Gd doped Bi$_2$Te$_3$ surface.** (a) STM image of pristine (Te-rich) Bi$_2$Te$_3$ surface measured at $V_S$ = +0.003 V. Native defects are denoted by A, B, and C. The defects A and B have clover-shaped protrusions, which are Te$_{Bi}$ anti-site defects residing on the first and the second Bi layers (Te$_{Bi1}$ and Te$_{Bi2}$), respectively. The defect C has clover-shaped depression, which is the Bi vacancy at the second Bi layer (V$_{Bi2}$). These defects were previously observed [2, 3]. The bright features of A and B defects in the empty-state STM image indicate that they act as donors, while the dark C defect might act as an acceptor. (b) STM image of Gd-doped Bi$_2$Te$_3$ surface measured at $V_S$ = +0.30 V. In addition to the defects A, B, and C observed in pristine Bi$_2$Te$_3$, the defects α and β are newly observed.

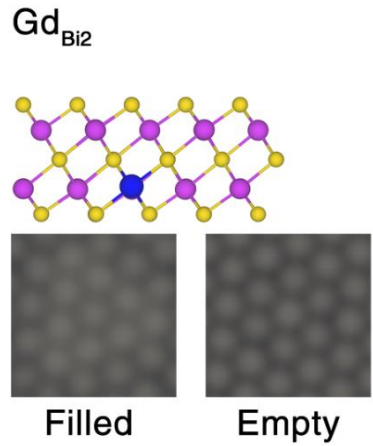

**Supplementary Figure 5. Simulated STM images of Gd$_{Bi2}$ defect.** The isolated Gd$_{Bi2}$ defect is indistinguishable in the simulated STM image, in stark contrast to the case of substitutional transition metal in the Bi2 layer [4, 5]. The filled- and empty-state images are simulated at $V_S$ = -0.6 and +0.6 V, respectively.

**Supplementary Table 1. Formation energy of considered defect structures.** The substitutional Gd at Bi sites ($Gd_{Bi1}$ and $Gd_{Bi2}$) are the most stable configurations. The next stable structure is the $Gd_{Bi1}$-$Bi_I$ pair.

| Defects | Formation energy (eV) | |
|---|---|---|
| | Te-rich condition | Bi-rich condition |
| $Gd_{ad}[top]$ | 7.30 | 7.30 |
| $Gd_{ad}[fcc]$ | 4.94 | 4.94 |
| $Gd_{ad}[hcp]$ | 4.84 | 4.84 |
| $Gd_I$ | 3.51 | 3.51 |
| $Gd_{Bi2}$-$Bi_I$ | 3.33 | 3.33 |
| $Gd_{Bi1}$ | 0.52 | 1.10 |
| $Gd_{Bi2}$ | 0.52 | 1.11 |
| $Gd_{Bi1}$-$Bi_{Te1}$ | 5.43 | 5.04 |
| $Gd_{Te1}$ | 5.49 | 5.10 |
| $Bi_{Te1}$ | 1.63 | 0.66 |
| $Bi_{Te3}$ | 1.42 | 0.45 |
| $Bi_{ad}$ | 2.07 | 1.49 |
| $Bi_I$ | 3.12 | 2.54 |
| $V_{Bi1}$ | 2.13 | 2.71 |
| $V_{Bi2}$ | 2.04 | 2.63 |

**Supplementary Note 1. Formation energy of Gd-related defects in $Bi_2Te_3$ surface.**

We performed the first-principles calculation to examine stable Gd-related defects in $Bi_2Te_3$ surface. Single Gd atom can be adsorbed on the surface ($Gd_{ad}$), intercalated in the van der Waals (vdW) gap ($Gd_I$), and substituted for Bi ($Gd_{Bi}$) and Te ($Gd_{Te}$) atoms. There are three distinct adsorption sites; on-top of Te atom (top), on-top of Bi atom (hcp), and hollow (fcc) sites. Likewise, Gd atom can occupy these three sites within the vdW gap. Supplementary Table 1 shows the formation energy ($E_f$) of all configurations calculated in this work.

The formation energy ($E_f$) was calculated by using the formula, $E_f = E[total] - E[Bi_2Te_3] - \sum \Delta n_i \cdot \mu_i$, where $E[total]$ is the total energy of the supercell containing defects and $E[Bi_2Te_3]$ is the total energy of the pristine $Bi_2Te_3$ surface. $\Delta n_i$ is the change of the number of atoms ($i$ = Gd, Bi, and Te) in the supercell. $\mu_i$ is the chemical potential of Gd, Bi, and Te. In the thermodynamic equilibrium for the stable growth of $Bi_2Te_3$, the chemical potentials of Bi and Te are not independent, $2\mu_{Bi} + 3\mu_{Te} = \Delta H_f(Bi_2Te_3)$, where $\Delta H$ is the formation enthalpy. We have $\mu_{Bi} = \mu_{Bi}^{bulk}$ in Bi-rich condition and $\mu_{Te} = \mu_{Te}^{bulk}$ in Te-rich condition. The possible range of chemical potentials are obtained, as follows. $\frac{1}{3}\Delta H_f(Bi_2Te_3) + \mu_{Te}^{bulk} \leq \mu_{Te} \leq \mu_{Te}^{bulk}$, $\frac{1}{2}\Delta H_f(Bi_2Te_3) + \mu_{Bi}^{bulk} \leq \mu_{Bi} \leq \mu_{Bi}^{bulk}$, and $\mu_{Gd} \leq \mu_{Gd}^{bulk}$. To avoid forming a Gd-Te compound, the Gd chemical potential is restricted: $\mu_{Gd} + 3\mu_{Te} \leq \Delta H_f(GdTe_3)$. The upper bound of $\mu_{Gd}$, $\mu_{Gd}^{max}$, depends on the crystal growth condition: $\mu_{Gd}^{max}$(Te-rich condition) $< \mu_{Gd}^{max}$(Bi-rich condition) $< \mu_{Gd}^{bulk}$. The lower bound of $\mu_{Gd}$ is taken to be negative infinity. Since the Gd chemical potential is permitted under Bi- and Te-rich condition, $\mu_{Gd}^{max} = \mu_{Gd}^{max}$(Te-rich condition).

**Supplementary Reference**